\documentclass[journal,12pt,onecolumn,draftclsnofoot]{IEEEtran}
\ifCLASSOPTIONcompsoc
  \usepackage[nocompress]{cite}
\else
  \usepackage{cite}
\fi
\ifCLASSINFOpdf
\else
\fi
\usepackage{amssymb}
\usepackage{amsmath}

\usepackage{array}

\usepackage[dvipsnames]{xcolor}	

\usepackage{pgfplots}
\usepackage{algorithm, algpseudocode}
\usepackage{enumerate}
\usepackage{graphicx}
\usepackage{bm}

\usetikzlibrary{quotes,arrows.meta}
\usetikzlibrary{positioning}
\usetikzlibrary{spy}
\usetikzlibrary{backgrounds}

\usetikzlibrary{3d} 

\begin{document}
	
	\title{Efficient Non-sequential Division for FPGAs}
	
	\author{Michael~Lunglmayr,~\IEEEmembership{Member,~IEEE}
		\thanks{Institute of Signal Processing, Johannes Kepler University Linz, Austria, corresponding e-mails: michael.lunglmayr@jku.at}
	}
	
	\maketitle
	
	\begin{abstract}
		The division operation is important for many areas of data processing. Especially considering today's demand for hardware accelerators for machine learning algorithms, there is a high demand for an efficient calculation of the division function, e.g. for averaging operations or the online calculation of activation functions. For such algorithms, which are often iterative in nature, one would like to have a non-sequential way of calculating the division operation. The work presents such an approach especially tailored to FPGAs as processing platforms. It is based on an efficient way of calculating the reciprocal operation, based on a low complexity approximation combined with a correction function. The described approach allows approximating the division operation (with errors that can be made arbitrarily low), within one clock cycle using only low hardware requirements.  
		These hardware requirements are scaleable depending on the desired precision. We show results obtained by synthesis and hardware simulations demonstrating the low complexity and high clock speeds achievable with the described method compared to other methods described in the literature.
	\end{abstract}
	
	\begin{IEEEkeywords}
		Division, Computation architecture, average calculation, digital hardware
	\end{IEEEkeywords}
	
	\IEEEpeerreviewmaketitle
	
	\section{Introduction}
	The division operation is considered to be the most complex of all four basic arithmetic operations \cite{meyer2004digital}. This complexity is mainly reflected in the effort required for calculating the reciprocal operation. If one has the reciprocal $1/x$ of a number $x$, the result of a division can be obtained by multiplication with such a reciprocal. 
	The use cases of efficient reciprocal operations are manifold. 
	Especially for iterative algorithms, where the division or reciprocal operation often has to be performed within an iteration, one would like to have a fast
	reciprocal operation. This is crucial for algorithms where the operand of the reciprocal operation depends on results calculated in previous iterations. Such dependencies often hinder an efficient pipelining of calculation units. Examples are algorithms where divisions are used to calculate stopping criteria \cite{stoppingcrit} or architectures using non-linear functions \cite{actfunc} based on divisions. In recent years, Field Programmable Gate Arrays (FPGA) have become increasingly popular as platforms for data science applications. Examples of works where FPGAs have been successfully used span from training support vector machines \cite{trainsvn}, over convolutional neural network (CNN) implementations \cite{fpgacnn}, to accelerators for recommendation algorithms \cite{recomenda}. FPGAs allow for massive parallelism combined with easy reconfigurability. Having an efficient approach to calculating  division is especially crucial when implementing algorithms on FPGAs where typically no dedicated hardware units are available for accelerating the division.  	
	Often hardware dividers are avoided or approximated in data processing architectures  \cite{stoppingcrit, actfunc, divAvoided} as ``division consumes too much cycles and area'' \cite{divAvoided} or are delegated to a connected CPU \cite{divDelegated}. 
	
	One of the reasons is that, typically, algorithms for division/reciprocal are sequential in nature. 
	Classical algorithms for implementing a division are often based on digit recurrence algorithms (e.g. see \cite{DivisionOverView} and the references therein), while algorithms for applications where speed is more important than precision often rely on iterative algorithms. Such algorithms are typically either based on Newton's method \cite{raps1,raps2,ScalingLessNewton} or on Goldschmidt's algorithm \cite{Gold1, Gold2, Goldschmidt}. Other approaches use lookup tables \cite{VLSICell} or combine lookup tables with iterative methods \cite{tab1, HighSpeedTC}. 
	In this work, we present an approach for a non-sequential division calculation, that can be calculated within one clock cycle (as we demonstrate in Sect.~\ref{sect:impl}; however, one can spend more clock cycles to increase the maximum clock frequency). Having a method allowing performing the division in only a few clock cycles (for the proposed method $1-5$) while still allowing for clock frequencies that are typical for large data processing architectures on FPGAs ($\sim100 - 300$ MHz) enables maximum flexibility when using the proposed method in a design. 
	
	The proposed method does not require Look-up tables and does not require any scaling of the input (as e.g. some implementations do that rely on Newton's method \cite{raps1}). 
	We will describe the approach that uses a simple approximation of the reciprocal function in combination with a correction function that is represented by a polynomial (where the degree of the polynomial defines the hardware requirements as well as the precision of the result). This correction function will, depending on the input value $x$ of the reciprocal, give a correction factor that is applied to the output of a more coarse approximation of the reciprocal. We will show that, with one correction function, the whole number range can be covered. In its simplest form, this will allow calculating the reciprocal operation using only three multiplications (in addition to other, less complex operations) with absolute errors below $10^{-3}$ for small input values $x$ and with errors smaller than $10^{-7}$ for high input values $x$. By increasing the number of multiplications, i.e. the degree of the correction polynomial, the precision can be further increased. 
	
	In this paper, first, the coarse reciprocal approximation is described. Then we derive the correction function and show how this correction function can be approximated by polynomials. We analyze the precision that can be achieved with the combined concept. We then present architectures that have been designed and optimized for correction polynomials of degrees $2$ and $4$ and finally show synthesis results demonstrating the low complexity and high clock speeds achievable with this method. We compare these results to state-of-the-art solutions demonstrating the gains achievable with this novel method.
	
	
	\section{Approximate Reciprocal calculation}
	In \cite{ApproxOOX}, we proposed a low-complexity approximation of the reciprocal function. We will repeat the main principle, as it is utilized for this approach, for the reader's convenience, and describe the approach such that the proposed method of this work can be easily built on.
	The idea of the approximation is the following. The function $1/x$ is linearized over the intervals $[2^z,2^{z+1})$ using $z = \lfloor log_2(x) \rfloor$ for each positive value $x$. For simplicity, we assume in this work that $x > 1$. Although the presented approach also works for positive values smaller than one (for negative values, one can convert to positive values first and then correct the sign of the output accordingly), the number ranges change depending on whether $x$ is smaller or larger than $1$. Assuming $x > 1$, the result conveniently fits into a fixed point format with only one bit in front of the comma, (the so-called fractional $1.(B-1)$ format of $B$-bits) that is often used in signal processing.  For easier readability, we will further assume that in each of the following equations $z$ is always $\lfloor \text{log}_2(x) \rfloor \in \mathbb{Z}$ of the $x$ appearing in the same context. 
	
\begin{figure}[h]
		\centering
		\includegraphics[width=0.95\columnwidth]{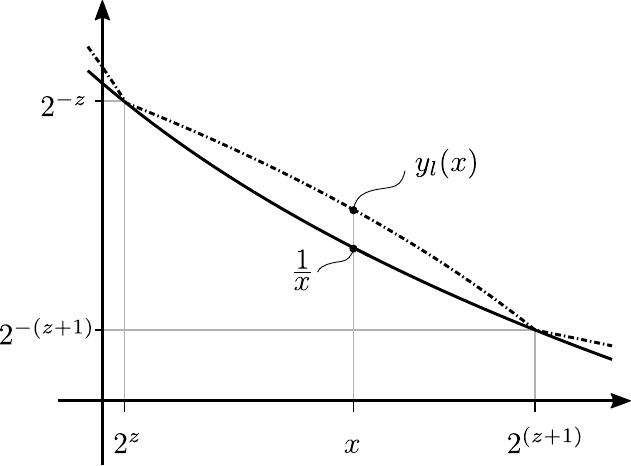}
		\caption{Piecewise approximation of $1/x$.}
		\label{fig:piecewise}
	\end{figure}

	Fig.~\ref{fig:piecewise} graphically shows the mentioned principle of the approximation $y_l(x)$.
	A value $y_l(x)$ between $[2^{-z},2^{-(z+1)})$ can be calculated using a linear combination factor $a \in [0,1)$
	as 
	\begin{align}
	  (1-a) 2^{-z} + a 2^{-(z+1)} = (2-a) 2^{-(z+1)}.
	\end{align}
	The value of $a$ corresponding to a value of $x \in [2^{-z},2^{-(z+1)}) $ can be calculated using the following considerations based on similar triangles.  $x \in [2^{-z},2^{-(z+1)}) $ can also be represented by a linear combination using $a \in [0,1)$:
	\begin{align}
		x = (1-a) 2^z + a 2^{z+1}.
		\label{eqn:xlin}
	\end{align}
	This allows calculating $a$ for a given value of $x$ as
	\begin{align}
		a = (x-2^z)/2^z = x2^{-z} - 1,
		\label{eqn:afromx}
	\end{align}
	and subsequently the corresponding value of $y_l(x)$:
	\begin{align}	
		y_l(x) = (3-x2^{-z})2^{-(z+1)}.
		\label{eqn:coarse}
	\end{align}
	The coarse approximation of (\ref{eqn:coarse}) can be implemented with very low complexity \cite{ApproxOOX}. To obtain $2^{-z}$, with $ z = \lfloor \text{log}_2(x) \rfloor$, one can use a leading one detector (to obtain $2^{z}$) followed by a bit reversal at the comma point. The multiplications with the powers of two in (\ref{eqn:coarse}) can then be realized via shift operations. 

	Investigating the relative error between $1/x$ and the approximation $y_l(x)$ showed that this error follows a curve of similar shape in each interval $[2^{z},2^{z+1})$. This led to the idea of correcting the approximation by multiplication with correction factors $\gamma(x)$ such that 
	\begin{align}
		\gamma(x) {(3-x 2^{-z}) 2^{-(z+1)}} = 1/x.
	\end{align}
	\section{Correction function}
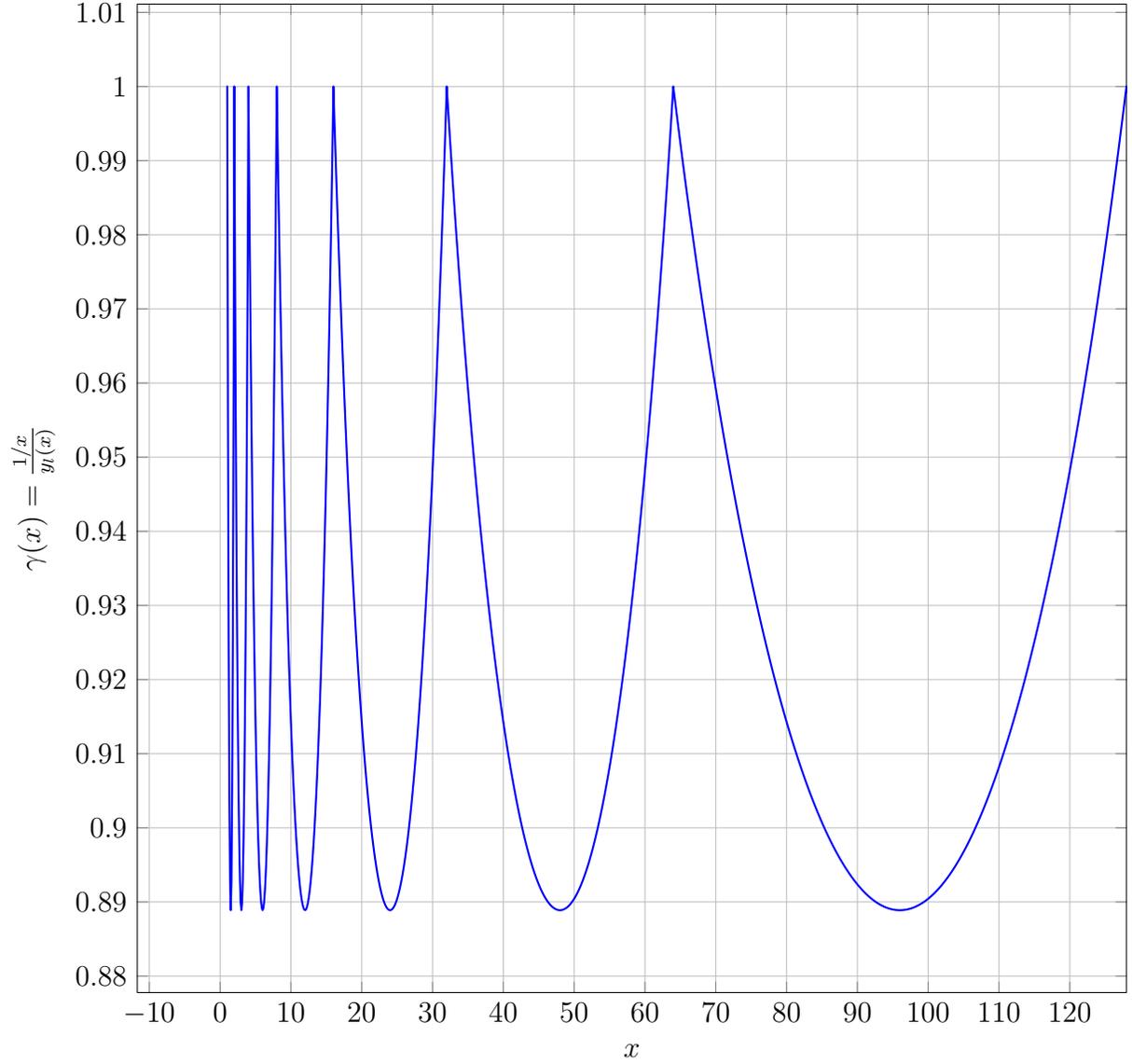
\begin{figure}[tb]
		\begin{center}
			\begin{tikzpicture}[line cap=round,line join=round]
			\begin{axis}[compat=newest, 
			width=.95\columnwidth, height =.95\columnwidth,log basis y=10, grid, xlabel=Iteration, 
			ylabel={ $\gamma(x) = \frac{1/x}{y_l(x)}$ }, 
			xlabel={ $x$ }, 
			xmax = 128,
			legend style={at={(1.0,1.0)},anchor=north east, font=\scriptsize},
			legend cell align=left,
			legend columns = {1},
			/pgf/number format/.cd, 1000 sep={}]
			\addplot[color=blue, thick, style=solid] table[x index = 0, y index = 1] {./correction_factor.dat};
			\end{axis}
			in node[fill=white] at (magnifyglass);
			\end{tikzpicture}
			\caption{Factors between $1/x$ and $y_l(x)$. \label{fig:correction_factor}}
		\end{center}
	\end{figure}

	Fig.~\ref{fig:correction_factor} graphically shows the correction factors over $x$.
	For an interval $[2^{z},2^{z+1})$, one can calculate such factors as giving the correction function
	\begin{align}
		\gamma(x) &= \frac{1}{x (3- x 2^{-z}) 2^{-(z+1)}} 
	\end{align}
	
	Using the abovementioned fact in \eqref{eqn:xlin}, that each number $x$ of an interval $[2^z, 2^{z+1})$ can be represented by a linear combination of the upper and lower interval
	one can formulate the correction function for $a \in [0,1]$ as
	\begin{align}
		\gamma(a) &= \frac{2}{3(1+a)-(1+a)^2}. \label{eqn:gamma} 
	\end{align}
	This formulation naturally stretches the correction function over each interval $[2^z, 2^{z+1})$. The value $a$ can be calculated from $x$ as it is shown in (\ref{eqn:afromx}). The most important part $x2^{-z}$ of (\ref{eqn:afromx}) already appears in the calculation of (\ref{eqn:coarse}) and can thus be reused from this calculation.
	
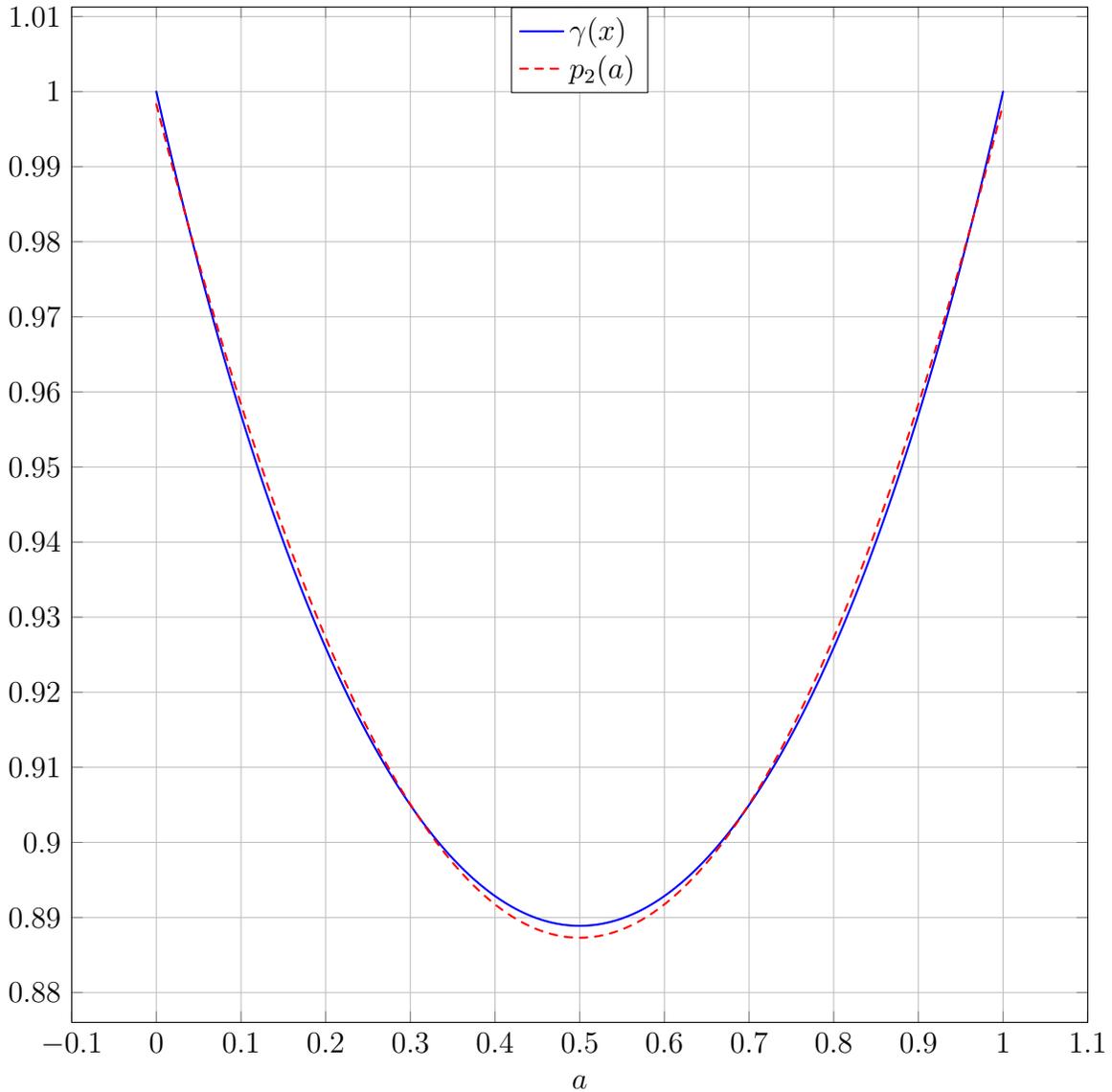
\begin{figure}[h]
		\centering
		\begin{tikzpicture}[line cap=round,line join=round]
		\begin{axis}[compat=newest, 
		width=.95\columnwidth, height =.95\columnwidth,log basis y=10, grid, xlabel=Iteration, 
		ylabel={ }, 
		xlabel={ $a$ }, 
		legend style={at={(.5,1.0)},anchor=north},
		legend cell align=left,
		legend columns = {1},
		/pgf/number format/.cd, 1000 sep={}]
		\addplot[color=blue, thick, style=solid] table[x index = 0, y index = 1] {./gvspoly2.dat};
		\addlegendentry{ $\gamma(x)$ }
		\addplot[color=red, thick, style=dashed] table[x index = 0, y index = 2] {./gvspoly2.dat};
		\addlegendentry{ $p_2(a)$ }
		\end{axis}
		\end{tikzpicture}
		\caption{Correction function and polynomial approximation of degree $2$.}
		\label{fig:CorrFun}
	\end{figure}

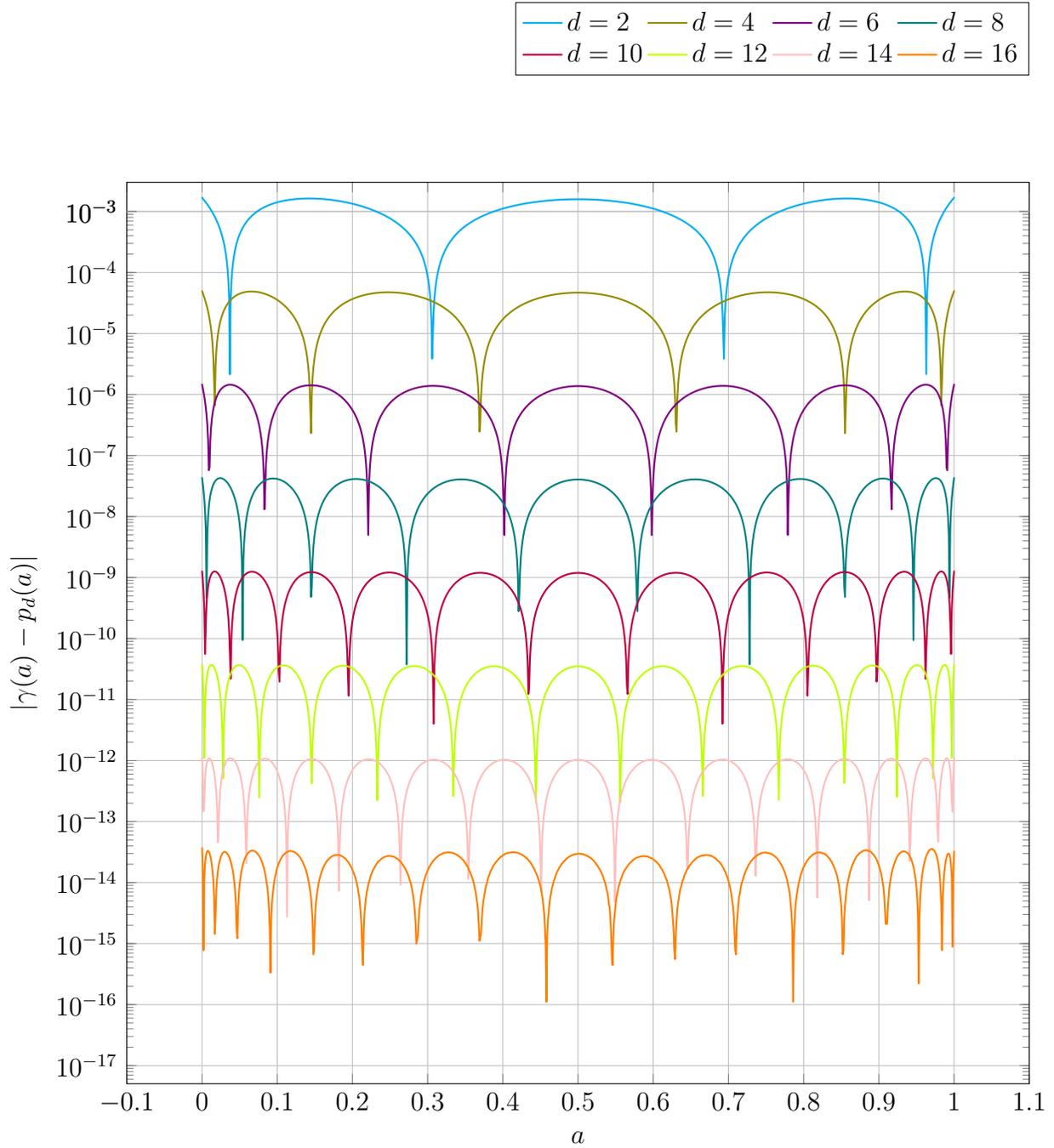
\begin{figure}[h]
		\centering
		\begin{tikzpicture}[line cap=round,line join=round]
		\begin{semilogyaxis}[compat=newest, 
		width=.95\columnwidth, height =.95\columnwidth,log basis y=10, grid, xlabel=Iteration, 
		ylabel={  $|\gamma(a)-p_d(a)|$ }, 
		xlabel={ $a$ }, 
		ymax = 3e-3,
		extra y ticks={1e-3},
		legend style={at={(1.0,1.2)},anchor=north east},
		legend cell align=left,
		legend columns = {4},
		/pgf/number format/.cd, 1000 sep={}]
		\addplot[color=cyan, thick, style=solid, unbounded coords=discard] table[x index = 0, y index = 1] {./ErrorsPolyGamma.dat};
		\addlegendentry{ $d=2$}
		\addplot[color=olive, thick, style=solid, unbounded coords=discard] table[x index = 0, y index = 2] {./ErrorsPolyGamma.dat};
		\addlegendentry{ $d=4$}
		\addplot[color=violet, thick, style=solid, unbounded coords=discard] table[x index = 0, y index = 3] {./ErrorsPolyGamma.dat};
		\addlegendentry{ $d=6$}
		\addplot[color=teal, thick, style=solid, unbounded coords=discard] table[x index = 0, y index = 4] {./ErrorsPolyGamma.dat};
		\addlegendentry{ $d=8$}
		\addplot[color=purple, thick, style=solid, unbounded coords=discard] table[x index = 0, y index = 5] {./ErrorsPolyGamma.dat};
		\addlegendentry{ $d=10$}
		\addplot[color=lime, thick, style=solid, unbounded coords=discard] table[x index = 0, y index = 6] {./ErrorsPolyGamma.dat};
		\addlegendentry{ $d=12$}
		\addplot[color=pink, thick, style=solid, unbounded coords=discard] table[x index = 0, y index = 7] {./ErrorsPolyGamma.dat};
		\addlegendentry{ $d=14$}
		\addplot[color=orange, thick, style=solid, unbounded coords=discard] table[x index = 0, y index = 8] {./ErrorsPolyGamma.dat};
		\addlegendentry{ $d=16$}
		\end{semilogyaxis}
		in node[fill=white] at (magnifyglass);
		\end{tikzpicture}
		\caption{Absolute approximation errors for polynomials of different degrees $d$}
		\label{fig:ApproxPrec}
	\end{figure}

	To obtain the values of $\gamma(a)$, one would require to calculate a reciprocal function, which would of course not be very helpful if one's aim is to use it in an approximation of the reciprocal. One can, however, approximate the correction function (\ref{eqn:gamma}) with polynomials. Fig.~\ref{fig:CorrFun} shows a plot of (\ref{eqn:gamma}) as well as a least squares (LS) fitted polynomial of degree $2$, $p_2(a)$. For polynomials $p_d(a)$ of higher degrees $d$, one would not see a difference to $\gamma(a)$ in this figure. 
	For this reason, we plotted the errors for polynomials of higher degrees ${|\gamma(a) - p_d(a)|}$ in Fig. ~\ref{fig:ApproxPrec}, using a logarithmic scale for the error values. Higher degrees than plotted in Fig.~\ref{fig:ApproxPrec} are limited by the precision of the double floating-point format. 
	
	The coefficients for all polynomials in this work have been obtained by sampling $a$ at Chebyshev nodes stretched to fit the interval $[0,1]$: 
	\begin{align}
		a = (\text{Re}(e^{\mathrm{i}\theta})+1)/2.
	\end{align}
	We used 
	$
	\theta \in [-\pi, -\pi+1\!\cdot\! 10^{-5}, -\pi+2\!\cdot\! 10^{-5}, \ldots, -1\!\cdot\! 10^{-5},0]	
	$\\
	\noindent
	in this work. The reason for using Chebyshev nodes for sampling the correction function is to reduce edge effects at the interval border \cite{ElementaryMuller}. By weighting the errors at the interval borders stronger, the approximation results in a relatively constant error over the whole interval $[0,1]$ (as can also be seen in Fig.~\ref{fig:ApproxPrec}). If one uniformly samples $a$ for the LS fit, the errors would increase at the interval borders $0$ and $1$, respectively. 
	Tab.~\ref{tab:polys} shows the coefficient vectors of the polynomials in the format $[c_d, c_{d-1}, c_0 ]$ describing the polynomials of the form 
	\begin{align}
		p_d(a) = \sum_{j=0}^{d} c_j a^j. 
	\end{align}
	\setlength{\extrarowheight}{.2cm}
	\begin{table*}
		
		\centering
		\caption{Polynomial coefficients of different degrees to approximate $\gamma(a)$. \label{tab:polys} }
		\begin{tabular}{|c|p{14cm}|}
			\hline
			degree	$d$ & coefficient vector  \\ 
			\hline 
			$2$ &  $[0.444059373310529,-0.444059378998574,0.998316470026731]$ \vspace{.2cm }\\
			\hline
			$4$ & $[0.209150199411479,-0.418300401501980,0.705497065458358,-0.496346863702732,0.999950441820227]$ \vspace{.2cm }\\
			\hline
			$6$ & $[0.098508912421565,-0.295526738526111,0.541617753193361,-0.590690940810007,0.745901070861286,$ $-0.499810057154873,0.999998541152684]$ \vspace{.2cm }\\
			\hline
			$8$ &$[0.046397306119941,-0.185589225073683,0.400091390713315,-0.550711883119918,0.661235760287096,$ $-0.621139145442145,0.749707092244658,-0.499991295729843,0.999999957055819]$ \vspace{.2cm }\\
			\hline
			$10$ & $[0.021852946554248,-0.109264732943323,0.278584857665575,-0.458751032417924,0.594292338298524,$ $-0.636160280553253,0.684123714654436,-0.624660366579099,0.749982187229885,-0.499999631909091,$ $0.999999998735847]$ \vspace{.2cm }\\
			\hline
			$12$ & $[0.010292651888006,-0.061755913969247,0.183962224976114,-0.353715232937807,0.512340641363993,$    $-0.606386252715474,0.656490500342412,-0.653408505661428,0.687155627762944,-0.624974786919717,$ $0.749999031146584,-0.499999985276385,0.999999999962789]$ \vspace{.2cm }\\
			\hline
			$14$ & $[0.004847557219273,-0.033932867266309,0.116332750240922,-0.256869471884100,0.418707931272722,$ $-0.547635332581819,0.624725982459027,-0.652257742453188,0.669535180214450,-0.655925894486105,$ $0.687470300538450,-0.624998345227464,0.749999951387920,-0.499999999433776,0.999999999998901]$ \vspace{.2cm }\\
			\hline
			$16$ & $[0.002280382320975,-0.018242765293630,0.070976706318807,-0.177591353506595,0.324764737187105,$ $-0.469933369923315,0.576458350792034,-0.633272693434323,0.658848078752916,-0.662148630845731,$ $0.671580922961134,-0.656218168654479,0.687497704369226,-0.624999898693680,0.749999997627261,$ $-0.499999999977700,0.999999999999963]$ \vspace{.2cm }\\
			\hline
		\end{tabular}
		
	\end{table*}
	
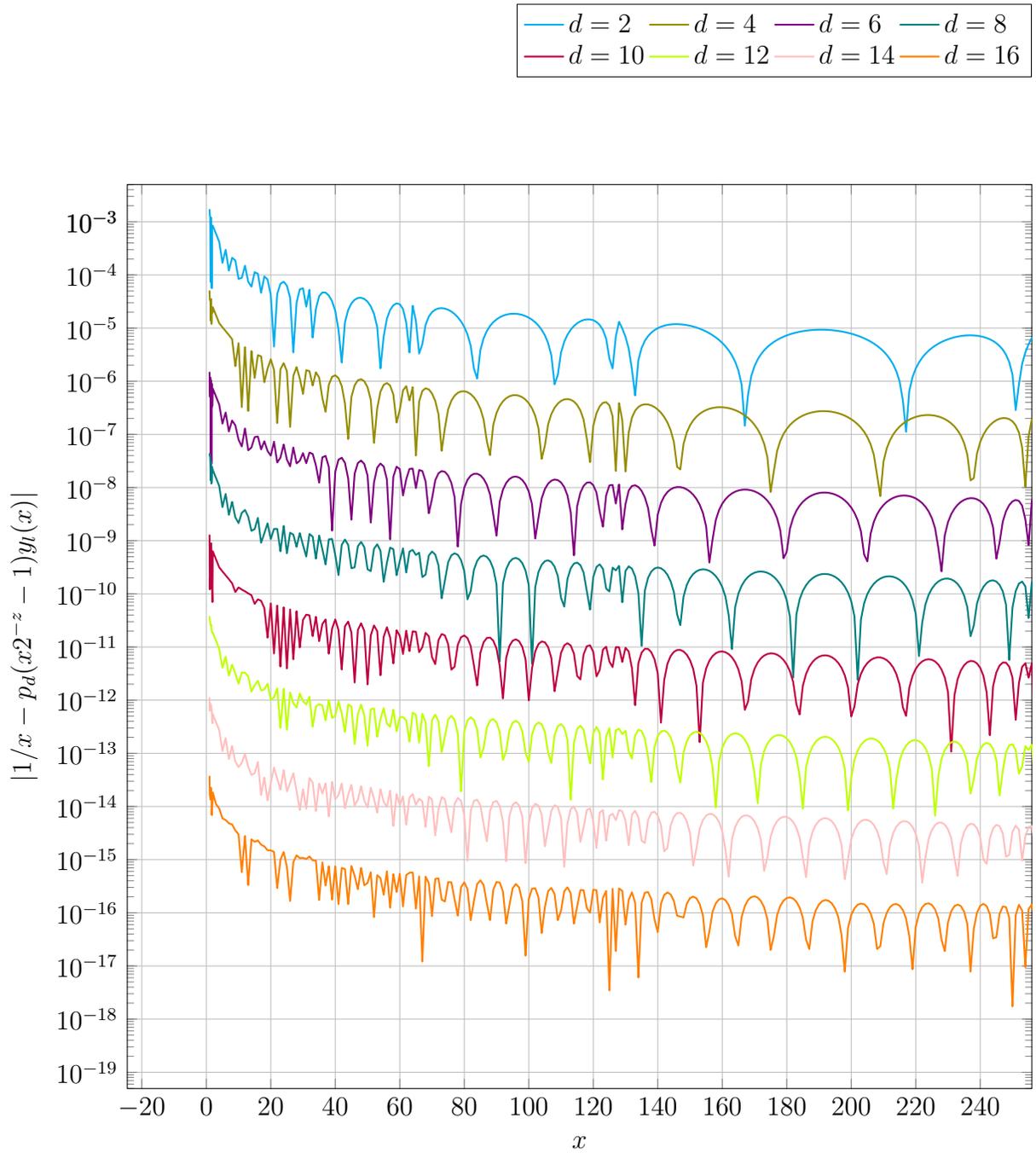
\begin{figure}[tb]
		\begin{center}
			\begin{tikzpicture}[line cap=round,line join=round]
			\begin{semilogyaxis}[compat=newest, 
			width=.95\columnwidth, height =.95\columnwidth,log basis y=10, grid, xlabel=Iteration, 
			ylabel={  $|{1/x}-p_d(x2^{-z}-1)y_l(x)|$ }, 
			xlabel={ $x$ }, 
			xmax = 256,
			ymax = 5e-3,
			extra y ticks={1e-3},
			legend style={at={(1.0,1.2)},anchor=north east},
			legend cell align=left,
			legend columns = {4},
			/pgf/number format/.cd, 1000 sep={}]
			\addplot[color=cyan, thick, style=solid, unbounded coords=discard] table[x index = 0, y index = 1] {./ErrorsPoly1x.dat};
			\addlegendentry{ $d=2$}
			\addplot[color=olive, thick, style=solid, unbounded coords=discard] table[x index = 0, y index = 2] {./ErrorsPoly1x.dat};
			\addlegendentry{ $d=4$}
			\addplot[color=violet, thick, style=solid, unbounded coords=discard] table[x index = 0, y index = 3] {./ErrorsPoly1x.dat};
			\addlegendentry{ $d=6$}
			\addplot[color=teal, thick, style=solid, unbounded coords=discard] table[x index = 0, y index = 4] {./ErrorsPoly1x.dat};
			\addlegendentry{ $d=8$}
			\addplot[color=purple, thick, style=solid, unbounded coords=discard] table[x index = 0, y index = 5] {./ErrorsPoly1x.dat};
			\addlegendentry{ $d=10$}
			\addplot[color=lime, thick, style=solid, unbounded coords=discard] table[x index = 0, y index = 6] {./ErrorsPoly1x.dat};
			\addlegendentry{ $d=12$}
			\addplot[color=pink, thick, style=solid, unbounded coords=discard] table[x index = 0, y index = 7] {./ErrorsPoly1x.dat};
			\addlegendentry{ $d=14$}
			\addplot[color=orange, thick, style=solid, unbounded coords=discard] table[x index = 0, y index = 8] {./ErrorsPoly1x.dat};
			\addlegendentry{ $d=16$}
			\end{semilogyaxis}
			in node[fill=white] at (magnifyglass);
			\end{tikzpicture}
			\caption{Absolute errors between $1/x$ and corrected approximation $y_l(x)$ using polynomials of different degrees $d$. \label{fig:errors_poly}}
		\end{center}
	\end{figure}

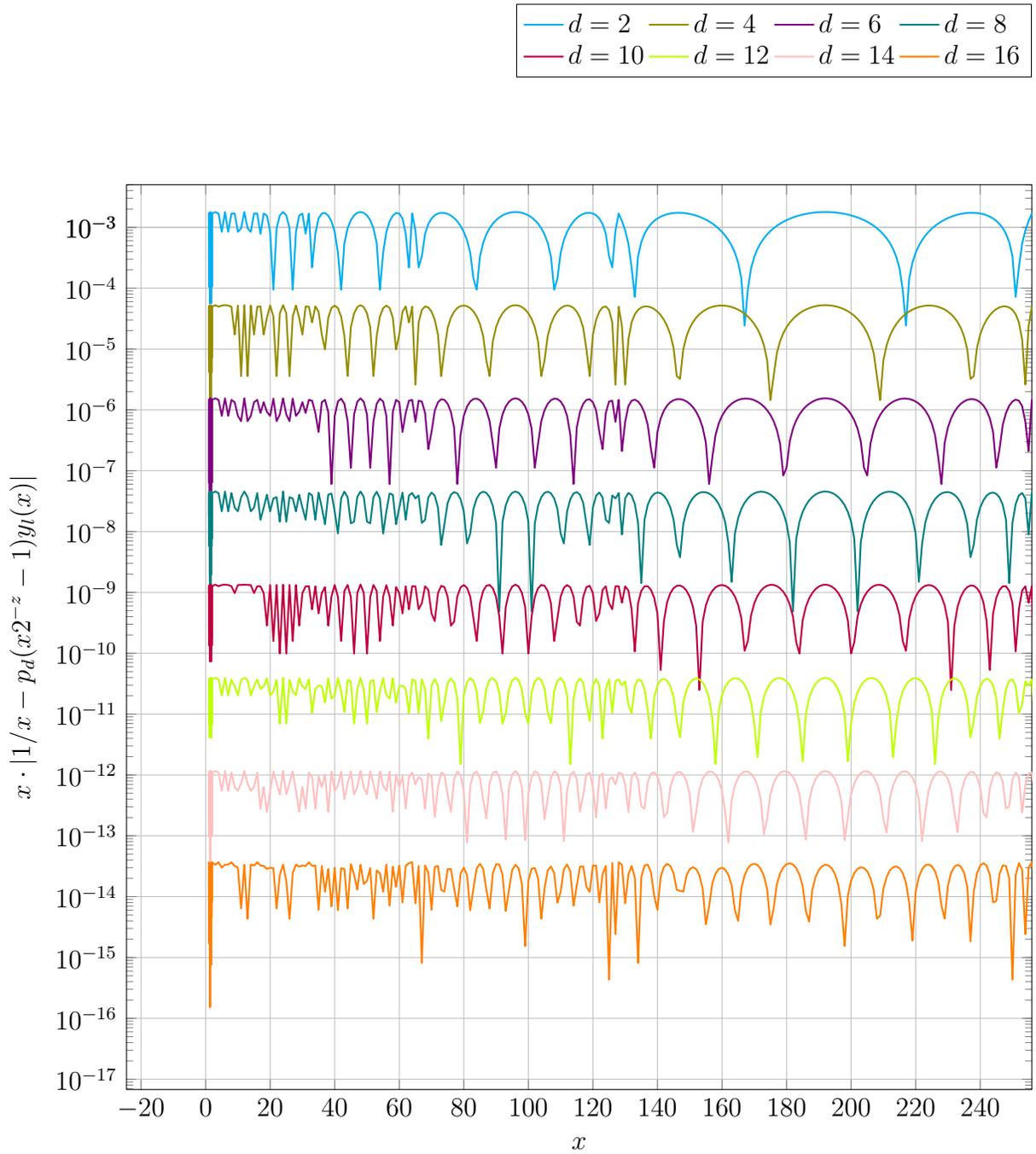
\begin{figure}[tb]
		\begin{center}
			\begin{tikzpicture}[line cap=round,line join=round]
			\begin{semilogyaxis}[compat=newest, 
			width=.95\columnwidth, height =.95\columnwidth,log basis y=10, grid, xlabel=Iteration, 
			ylabel={  $x \cdot |{1/x}-p_d(x2^{-z}-1)y_l(x)|$ }, 
			xlabel={ $x$ }, 
			xmax = 256,
			ymax = 5e-3,
			extra y ticks={1e-3},
			legend style={at={(1.0,1.2)},anchor=north east},
			legend cell align=left,
			legend columns = {4},
			/pgf/number format/.cd, 1000 sep={}]
			\addplot[color=cyan, thick, style=solid, unbounded coords=discard] table[x index = 0, y index = 1] {./RErrorsPoly1x.dat};
			\addlegendentry{ $d=2$}
			\addplot[color=olive, thick, style=solid, unbounded coords=discard] table[x index = 0, y index = 2] {./RErrorsPoly1x.dat};
			\addlegendentry{ $d=4$}
			\addplot[color=violet, thick, style=solid, unbounded coords=discard] table[x index = 0, y index = 3] {./RErrorsPoly1x.dat};
			\addlegendentry{ $d=6$}
			\addplot[color=teal, thick, style=solid, unbounded coords=discard] table[x index = 0, y index = 4] {./RErrorsPoly1x.dat};
			\addlegendentry{ $d=8$}
			\addplot[color=purple, thick, style=solid, unbounded coords=discard] table[x index = 0, y index = 5] {./RErrorsPoly1x.dat};
			\addlegendentry{ $d=10$}
			\addplot[color=lime, thick, style=solid, unbounded coords=discard] table[x index = 0, y index = 6] {./RErrorsPoly1x.dat};
			\addlegendentry{ $d=12$}
			\addplot[color=pink, thick, style=solid, unbounded coords=discard] table[x index = 0, y index = 7] {./RErrorsPoly1x.dat};
			\addlegendentry{ $d=14$}
			\addplot[color=orange, thick, style=solid, unbounded coords=discard] table[x index = 0, y index = 8] {./RErrorsPoly1x.dat};
			\addlegendentry{ $d=16$}
			\end{semilogyaxis}
			in node[fill=white] at (magnifyglass);
			\end{tikzpicture}
			\caption{Relative errors between $1/x$ and corrected approximation $y_l(x)$ using polynomials of different degrees $d$. \label{fig:Rerrors_poly}}
		\end{center}
	\end{figure}

	Using these correction polynomials, one can evaluate the 
	errors (compared to $1/x$) when being applied to the approximation $y_l(x)$. Such errors are plotted in Fig.~\ref{fig:errors_poly}.
	As one can see from this figure, even with a correction polynomial of degree $2$ for $x$ values larger than $1.6$ one is always below an absolute error of $10^{-3}$ (the maximum absolute error at $x>1$ is $1.684e-3$). Each degree increase by $2$, reduces the error by approximately a factor of $35$. 
	The relative error can be obtained by multiplying the error of Fig.~\ref{fig:errors_poly} with $x$ (it then gives the error related to $1/x$). It is shown in Fig.~\ref{fig:Rerrors_poly} for different degrees of the polynomials. Here, one can see that, for a polynomial of a given degree, the relative error has about the same magnitude for all $x$ values (although for better visibility, we only plotted $x$ values up to $256$, the same levels can be observed for larger values as well). 
	
	\section{Hardware Architectures}
	For the author's applications, the correction polynomials of degrees $2$ and $4$ are of main interest. Considering the error curves described in the sections before, these correction polynomials allow calculating the reciprocal with approximately $10$ bits precision and better when using the degree $2$ polynomial and approximately $16$ bits precision when using the degree $4$ polynomial (as described above, the absolute precision improves for large values of $x$). 
	
\begin{figure*}[t]
		\centering
		\includegraphics[width=1.1\columnwidth]{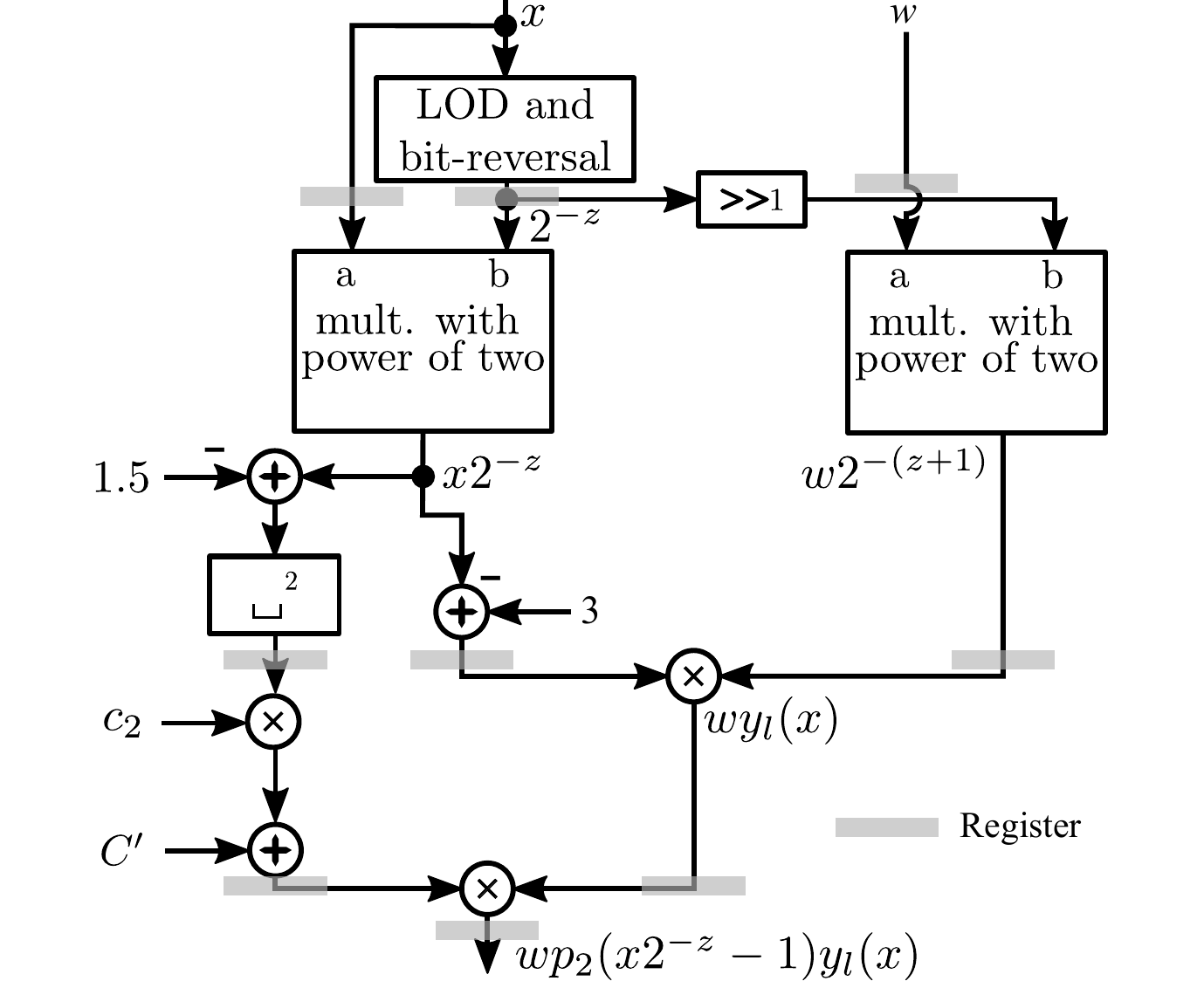}
		\caption{Architecture using correction polynomial of degree $2$.}
		\label{fig:poly2arc}
	\end{figure*}

	For these two polynomials, the architectures described in this section have been developed. If one has a look at the coefficients of the polynomial of degree $2$, one can see that the coefficients $c_2$ and $c_1$ are nearly equal in absolute value.
	Assuming the values to be equal (the difference in magnitude is of order $10^{-9}$) one can reformulate $p_2(a)$ as 
	\begin{align}
		p_2(a) = c_2 (a-0.5)^2 + c_0 - 0.25c_2,
		\label{eqn:p2}
	\end{align}
	where the constant can be combined into $C' = c_0 - 0.25c_2$.
	This allows incorporating this correction polynomial efficiently in the approximate $1/x$ architecture of \cite{ApproxOOX}, as it is schematically shown in Fig.~\ref{fig:poly2arc}. Here the input $a = x 2^{-z}-1$ of the polynomial is already combined with the subtraction of $0.5$ of (\ref{eqn:p2}), resulting in the subtraction with $1.5$ before the squaring operation.
	
	In this architecture, we already incorporated a multiplication of a dividend input $w$. Instead of calculating 
	\begin{align}
		w \cdot \left ( p_d(x 2^{-z} - 1 ) (3-x 2^{-z}) 2^{-(z+1)} \right ),
	\end{align}
	one can do the multiplication $w 2^{-(z+1)}$ within the architecture. This can reduce the bit length for the multiplier by combining the now shifted $w$ value with the rest of the division operation, especially for applications where $w$ has a large number of bits in front of the comma (that can be reduced after the right shift by $z+1$). An example application where this is convenient is when using the architecture for calculating averages over $N$ fractional ($1.B-1$) numbers. In this scenario, $x=N$ and, because $w$ is a sum of fractional numbers, $w 2^{-(z+1)}$ is then also a fractional  number.
	
\begin{figure*}[t]
		\centering
		\includegraphics[width=1.1\columnwidth]{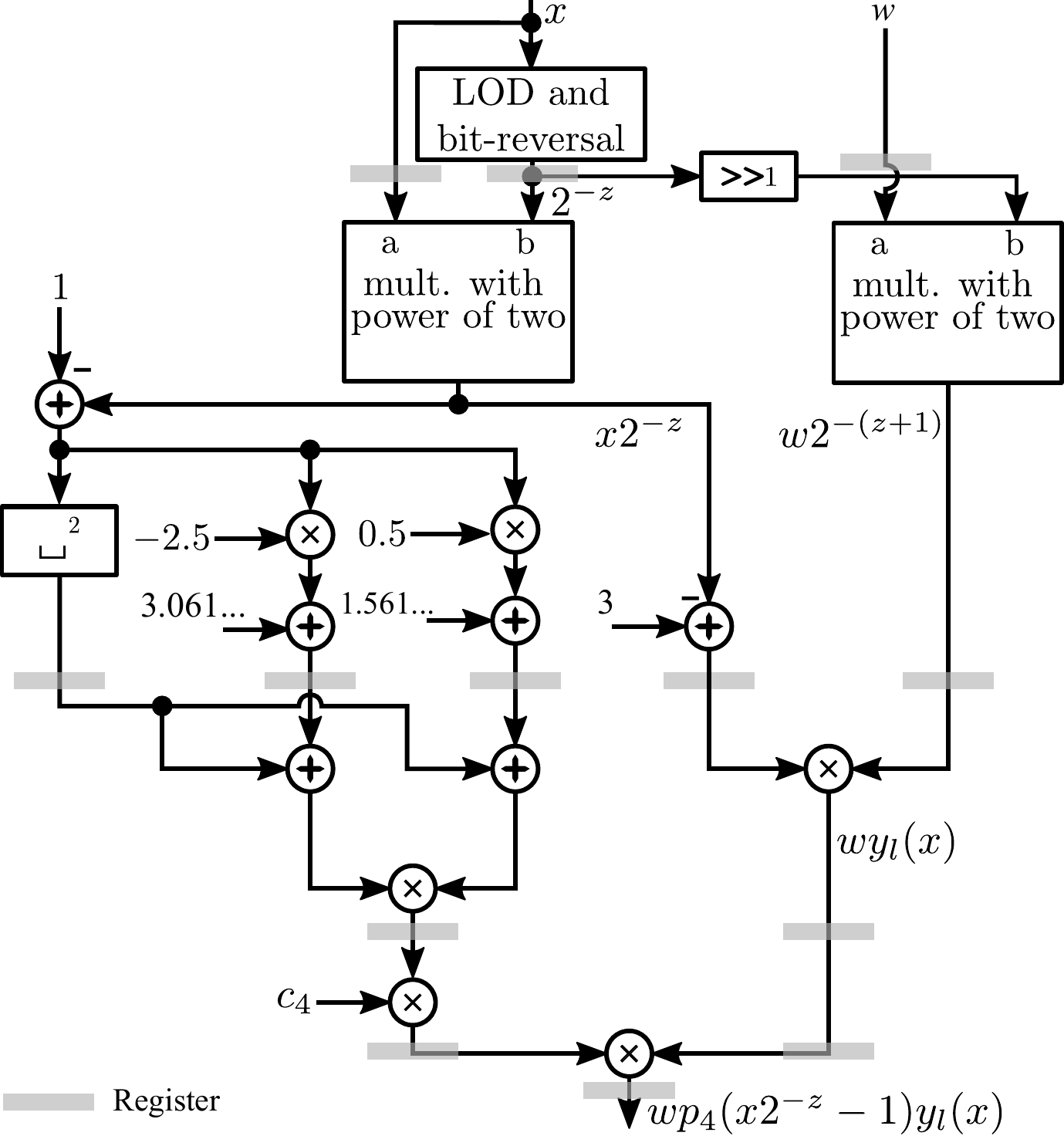}
		\caption{Architecture using correction polynomial of degree $4$.}
		\label{fig:poly4arc}
	\end{figure*}

	The polynomial of degree $4$ of Tab.~\ref{tab:polys} can be factored into the following form:
	\begin{align}
		&0.209150199411479 \;\cdot \nonumber \\ \nonumber &(3.0616168632399 - 2.500018461800448 a + 
		a^2) \;\cdot \\  &(1.561598389171924 + 0.5000184489913662 a + a^2)
	\end{align}
	In this form, the coefficients in front of $a$ can be rounded to $-2.5$ and $0.5$  without a noticeable effect on the approximation error (at least not for the synthesized bit lengths of this work).
	This allows performing the multiplications with these values using
	two shifts and one
	add operation plus a single shift operation.
	The architecture using these simplifications for the polynomial of degree $4$ is shown in Fig.~\ref{fig:poly4arc}.

	In Fig.~\ref{fig:poly2arc} and Fig.~\ref{fig:poly4arc}, the gray rectangles specify the register positions for the implementations using 
	$4 - 5$ clock cycles for calculating the result, as is described in the next section. For the one-clock-cycle variants, only the registers before the outputs were kept, all other registers have been removed.
	
		\begin{figure}[h]
		\begin{center}
			\begin{tikzpicture}[line cap=round,line join=round]
			\begin{semilogyaxis}[compat=newest, 
			width=.95\columnwidth, height =.95\columnwidth,log basis y=10, grid, xlabel=Iteration, 
			ylabel={  $|{1/x}-p_d(x2^{-z}-1)y_l(x)|$ }, 
			xlabel={ $x$ }, 
			xmax = 256,
			ymax = 1e-3,
			ymin = 1e-7,
			extra y ticks={1e-3},
			legend style={at={(1.0,1.0)},anchor=north east},
			legend cell align=left,
			legend columns = {1},
			/pgf/number format/.cd, 1000 sep={}]
			\addplot[color=cyan, thick, style=solid, unbounded coords=discard] table[x index = 0, y index = 1] {./e_abs_impl.dat};
			\addlegendentry{ $d=2$}
			\addplot[color=olive, thick, style=solid, unbounded coords=discard] table[x index = 0, y index = 2] {./e_abs_impl.dat};
			\addlegendentry{ $d=4$}
			\addplot[color=violet, thick, style=dashed, unbounded coords=discard] table[x index = 0, y index = 3] {./e_abs_impl.dat};
			\addlegendentry{ optimal $16$-bit}
			
			\coordinate (pt) at (axis cs:-3,7e-4);
			\coordinate (ptback) at (axis cs:27,.9e-3);
			\coordinate (ptcut) at (axis cs:0,2e-6);
			
			\end{semilogyaxis}
			%
			
			\end{tikzpicture}
			\caption{Absolute errors between $1/x$ and hardware implementations of degrees $2$ and $4$. \label{fig:e_abs_impl}}
		\end{center}
	\end{figure}
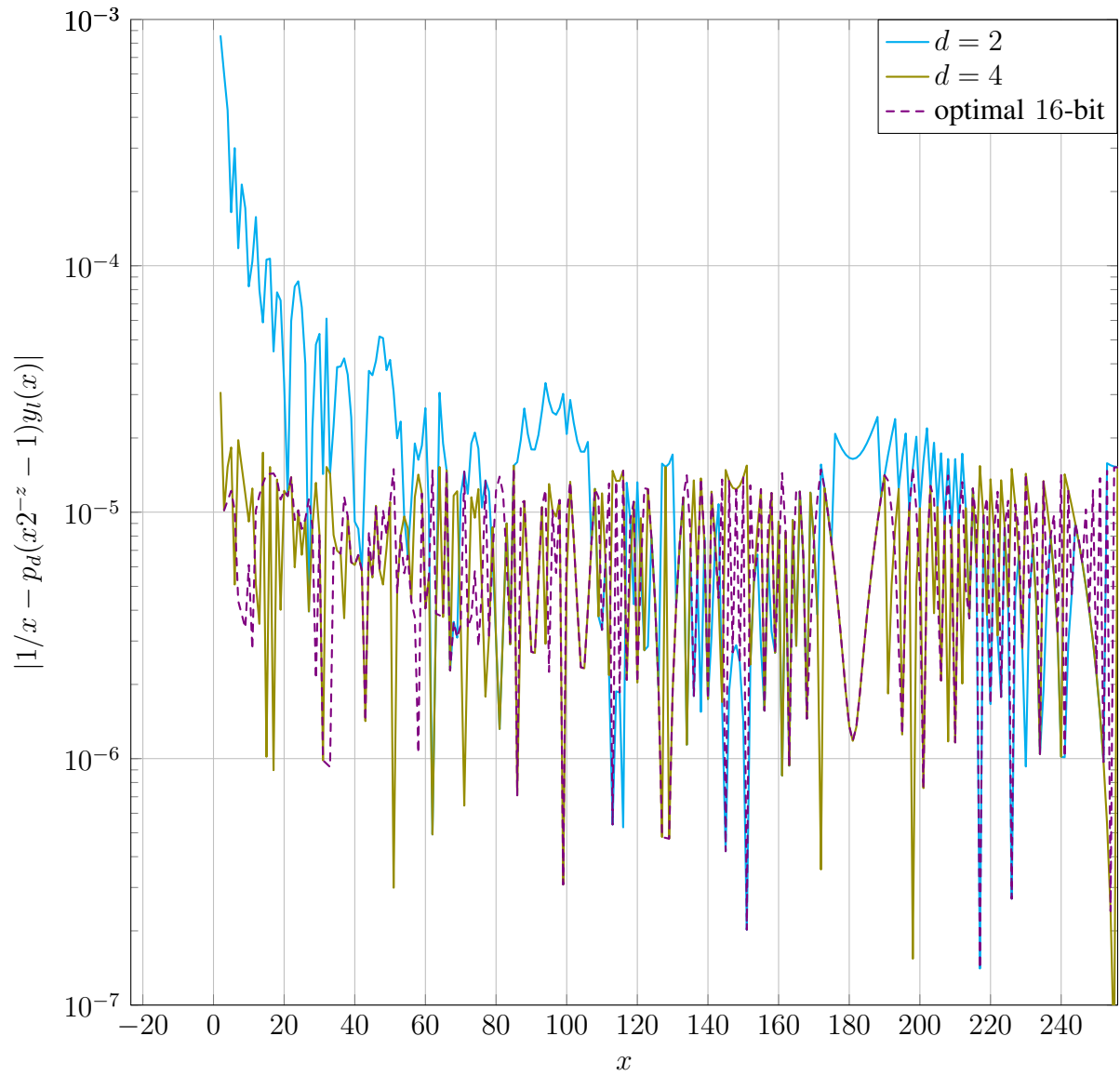
	
	\section{Implementation results}
	\label{sect:impl}
	Fig.~\ref{fig:e_abs_impl} shows absolute errors obtained by VHDL simulations of the implemented architecture, for the architectures of Fig.~\ref{fig:poly2arc} and Fig.~\ref{fig:poly4arc}, respectively. The bit lengths for this architecture have been chosen for the application scenario of calculating the average of up to $2^{15}$ $16$-bit fractional values, i.e., $x$ was chosen to have only $16$ bits in front of the comma, and $w$ had $16$ bits before and $15$ bits after the comma. Tab.~\ref{tab:synth} shows the synthesis results for this scenario using {Quartus Prime 19.1}. All internal calculations of the architectures have been performed with $17$ bits after the comma. The results shown in Fig.~\ref{fig:e_abs_impl}, have been obtained for $w=1$. For reference, we plotted the results when calculating $1/x$ in double precision rounded to the next $16$-bit fractional value (optimal $16$-bit). 
	As one can see from Fig.~\ref{fig:e_abs_impl}, the architecture for the polynomial of degree $4$ practically achieves a comparable precision to the optimal $16$-bit solution (except for $x$ values up to $4$, where the absolute error is slightly larger). For the degree $2$ architecture, one can achieve about $10$-bit precision for low $x$ values and up to $16$-bits precision for large $x$ values.
	Using larger internal calculation bit lengths can even increase the precision for large $x$ values, as the results of Fig.~\ref{fig:errors_poly} demonstrate.
	
	In Tab.~\ref{tab:synth}, we show synthesis results for two architectures of Fig.~\ref{fig:poly2arc} and Fig.~\ref{fig:poly4arc} for the IntelFGPA Stratix V (5SGSMD5K2F40C2), respectively. 
	\begin{table}[h]
		\centering
		\caption{Synthesis results for {Stratix V: 5SGSMD5K2F40C2} \label{tab:synth}}
		\begin{tabular}{|l|p{.6cm}|p{.6cm}|p{.6cm}|p{.6cm}|}
			\hline
			& \multicolumn{2}{c|}{degree $2$}                        & \multicolumn{2}{c|}{degree $4$} \\ \hline
			Clock cycles              & {$4$}        &$1$      & $5$          & $1$ \\ \hline
			ALMs (of $172,600$)       &  167        & 170      & 227          & 210    \\
			Registers (of $706,560$)  &  157        & 66       & 215          & 70     \\
			DSP blocks (of $1,590$)   &  4          & 4        & 5            & 5      \\
			Fmax Slow 900mV 85C (MHz) &  334.78     & 116.78   & 251          &  78.77 \\ \hline
		\end{tabular}
	\end{table}
	When using all intermediate registers as drawn in the figures, the architecture of Fig.~\ref{fig:poly2arc} requires $4$ clock cycles for calculation and the architecture of Fig.~\ref{fig:poly4arc} requires $5$ clock cycles. This might be an option when one can utilize the pipelining capabilities of the architectures. For iterative algorithms, where the input of a division unit often depends on values that are only known after completing an iteration one might want to reduce the number of clock cycles. To show the results for the most extreme reduction of registers in the architectures, the synthesis results are shown for the architectures without intermediate registers ($1$ clock cycle) as well. The obtain correct timing results, the inputs of the architectures have been registered as well (this accounts for the higher number of registers in the synthesis results). However, when using the described architecture as part of a larger architecture, such registers will be typically present in the entity feeding the inputs of the division unit. As one can see from these results, the maximum clock frequency drops by about a factor of $3$ for the one-clock-cycle architectures, compared to when using $4-5$ clock cycles. For single clock cycle implementations, one obtains a maximum clock frequency higher than $110$ MHz for degree $2$ and a maximum clock frequency of $78.77$ MHz for degree $4$. The clock frequencies of the $4-5$ clock cycle variant, according to the author's experience, are higher than the clock frequency than can be typically achieved for advanced data processing architectures (e.g. machine learning algorithms) when using this FPGA. 
	
	To put the synthesis results of Tab.~\ref{tab:synth} into perspective we synthesized the division operation using operators of the sfixed data type \cite{SFIXED} of VHDL. Tab.~\ref{tab:sfixed} shows the synthesis results when using these operators. We synthesized two approaches. The first, using the division operator calculating $w/x$ and the second multiplying $w$ with the \emph{reciprocal} function from sfixed called on $x$. The bit sizes of $w$ and $x$ have been the same as described above. The inputs and outputs have been again registered to obtain meaningful timing results.
		\begin{table}[h]
		\centering
		\caption{Synthesis results for {Stratix V: 5SGSMD5K2F40C2} using the functions of the sfixed \cite{SFIXED} library
			for comparison to the single clock implementations \label{tab:sfixed}}
		\begin{tabular}{|l|p{.6cm}|p{2cm}|}
			\hline
									&  $w/x$        &  $w*\text{reciprocal}(x)$ \\ \hline
			Clock cycles              & {$1$}       &$1$      \\ \hline
			ALMs (of $172,600$)       &  $970$        & $292$      \\
			Registers (of $706,560$)  &  $63$         & $32^\dagger$      \\
			DSP blocks (of $1,590$)   &  $0$          & $1$        \\
			Fmax Slow 900mV 85C (MHz) &  $15.18$    & $33.91$   \\ \hline
		\end{tabular}\\
	\vspace{.1cm}
	$^\dagger$ the other registers have been used from the DSP block.
	\end{table}
	As one can see in this table, when implementing the division in one clock cycle using the functions from the sfixed data type (leading to fully combinatorial implementations of the division) one obtains significantly slower designs than the designs proposed in this work. Even when including an on-board multiplier (the DSP-block) of the FPGA, the maximum clock speed is less than half when using the approach proposed in this work with a correction polynomial of degree $4$ in the single clock cycle configuration.
	
	\section{Comparison with State-of-the-Art}
	In Tab.~\ref{tab:compari}, we collected three synthesis results of efficient reciprocal FPGA implementations reported in the literature (achieving $16$-bit precision). In \cite{raps1}, a fixed point divider utilizing a reciprocal operation based on the Newton-Raphson algorithm is used followed by a multiplication operation. It uses a piecewise polynomial approximation for providing the seed of the Newton-Raphson algorithm. 
	 The comparison can be performed most fairly between \cite{raps1} and this work as comparable FPGAs from the same family are used. As Tab.~\ref{tab:compari} shows, the design of \cite{raps1} achieves a slightly smaller clock frequency than the proposed design of this work in the single clock cycle implementation but requires $3$ clock cycles instead.
	 
	 In \cite{ScalingLessNewton}, a design for a reciprocal operation is described. One can again implement a division operation using a multiplication after calculating the reciprocal. 
	 The design of \cite{ScalingLessNewton} is also based on the Newton-Raphson algorithm but uses pipelining to achieve higher clock speeds. While an impressive $740$ MHz clock speed is reported, the implementations of \cite{ScalingLessNewton} require $20$ and more clock cycles to calculate the reciprocal operation. The used VIRTEX-7 FPGA is typically rated for higher clock speeds than the Stratix V used in this work, and thus should give faster results. But even when directly comparing the delay times using the different FPGAs, our implementation is still twice as fast.
	\begin{table}[h]
		\centering
		\caption{ Literature reports on efficient reciprocal implementations \label{tab:compari}}
		\setlength\tabcolsep{3pt} 
		\setlength\extrarowheight{6pt}
		\begin{tabular}{|p{2.7cm}|p{1.5cm}|p{1.5cm}|p{1.5cm}|}
			\hline
			Work reported in        & \cite{ScalingLessNewton}			& \cite{ScalingLessNewton}	& \cite{raps1} \\
			\hline
			FPGA                    & VIRTEX-4 SX35     & VIRTEX-7 690T & Stratix-V 5SGXMA7 \\
			\hline
			Slices                  & 347               & not reported  & 339   \\
			LUT                     & 372               & 111           &       \\
			Registers               & 568               & 240           & 73    \\
			DSP blocks              & 7                 & 6-10          & 5     \\
			Clock cycles            & 25                & 20-31         & 3     \\
			Clock frequency         & 294.1             & 740           & 68.62 \\
			\hline
		\end{tabular}
	\end{table}
	When comparing the described architecture with these results, one can see that comparable FPGA utilization is achieved (the author again wants to point out that the comparison algorithms are only for calculating the reciprocal while the architecture of this work already includes the multiplication with the dividend $w$). Comparing the speed, one can see that the proposed method leads to up to $3$ times faster implementations (when comparing with the degree $4$ variant of this work) than the state-of-the-art methods. 
	
	\section{Conclusion}
	We presented a concept for efficient non-sequential division especially tailored for FPGA implementation. It is based on an efficient way of approximating the reciprocal operation by piece-wise linearization with node points as powers of two. Using a correction function, that can be approximated with a single polynomial for the whole number range allows implementation of the division with high precision and low hardware requirements even in a single clock cycle at high clock frequencies.

	\bibliographystyle{IEEEtran}
	\bibliography{mybib}

\begin{thebibliography}{10}
\providecommand{\url}[1]{#1}
\csname url@samestyle\endcsname
\providecommand{\newblock}{\relax}
\providecommand{\bibinfo}[2]{#2}
\providecommand{\BIBentrySTDinterwordspacing}{\spaceskip=0pt\relax}
\providecommand{\BIBentryALTinterwordstretchfactor}{4}
\providecommand{\BIBentryALTinterwordspacing}{\spaceskip=\fontdimen2\font plus
\BIBentryALTinterwordstretchfactor\fontdimen3\font minus
  \fontdimen4\font\relax}
\providecommand{\BIBforeignlanguage}[2]{{%
\expandafter\ifx\csname l@#1\endcsname\relax
\typeout{** WARNING: IEEEtran.bst: No hyphenation pattern has been}%
\typeout{** loaded for the language `#1'. Using the pattern for}%
\typeout{** the default language instead.}%
\else
\language=\csname l@#1\endcsname
\fi
#2}}
\providecommand{\BIBdecl}{\relax}
\BIBdecl

\bibitem{meyer2004digital}
U.~Meyer-Baese, \emph{{Digital signal processing with field programmable gate
  arrays}}.\hskip 1em plus 0.5em minus 0.4em\relax Springer Verlag, 2004.

\bibitem{stoppingcrit}
A.~Ibrahim and M.~Valle, ``{Real-Time Embedded Machine Learning for Tensorial
  Tactile Data Processing},'' \emph{IEEE Transactions on Circuits and Systems
  I: Regular Papers}, vol.~65, no.~11, pp. 3897--3906, 2018.

\bibitem{actfunc}
S.~Zheng, P.~Ouyang, D.~Song, X.~Li, L.~Liu, S.~Wei, and S.~Yin, ``{An
  Ultra-Low Power Binarized Convolutional Neural Network-Based Speech
  Recognition Processor With On-Chip Self-Learning},'' \emph{IEEE Transactions
  on Circuits and Systems I: Regular Papers}, vol.~66, no.~12, pp. 4648--4661,
  2019.

\bibitem{trainsvn}
J.~Dass, Y.~Narawane, R.~N. Mahapatra, and V.~Sarin, ``{Distributed Training of
  Support Vector Machine on a Multiple-FPGA System},'' \emph{IEEE Transactions
  on Computers}, vol.~69, no.~7, pp. 1015--1026, 2020.

\bibitem{fpgacnn}
R.~Neris, A.~Rodr\'iguez, R.~Guerra, S.~L\'opez, and R.~Sarmiento,
  ``{FPGA-Based Implementation of a CNN Architecture for the On-Board
  Processing of Very High-Resolution Remote Sensing Images},'' \emph{IEEE
  Journal of Selected Topics in Applied Earth Observations and Remote Sensing},
  vol.~15, pp. 3740--3750, 2022.

\bibitem{recomenda}
C.~Wang, L.~Gong, X.~Ma, X.~Li, and X.~Zhou, ``{WooKong: A Ubiquitous
  Accelerator for Recommendation Algorithms With Custom Instruction Sets on
  FPGA},'' \emph{IEEE Transactions on Computers}, vol.~69, no.~7, pp.
  1071--1082, 2020.

\bibitem{divAvoided}
G.~C. Cardarilli, L.~D. Nunzio, R.~Fazzolari, M.~Panella, M.~Re, A.~Rosato, and
  S.~Span, ``{A Parallel Hardware Implementation for 2-D Hierarchical
  Clustering Based on Fuzzy Logic},'' \emph{IEEE Transactions on Circuits and
  Systems II: Express Briefs}, vol.~68, no.~4, pp. 1428--1432, 2021.

\bibitem{divDelegated}
L.~Du, Y.~Du, and M.-C.~F. Chang, ``{A Reconfigurable 64-Dimension K-Means
  Clustering Accelerator With Adaptive Overflow Control},'' \emph{IEEE
  Transactions on Circuits and Systems II: Express Briefs}, vol.~67, no.~4, pp.
  760--764, 2020.

\bibitem{DivisionOverView}
S.~F. {Obermann} and M.~J. {Flynn}, ``{Division algorithms and
  implementations},'' \emph{IEEE Transactions on Computers}, vol.~46, no.~8,
  pp. 833--854, Aug 1997.

\bibitem{raps1}
A.~{Rodriguez-Garcia}, L.~{Pizano-Escalante}, R.~{Parra-Michel},
  O.~{Longoria-Gandara}, and J.~{Cortez}, ``{Fast fixed-point divider based on
  Newton-Raphson method and piecewise polynomial approximation},'' in
  \emph{2013 International Conference on Reconfigurable Computing and FPGAs
  (ReConFig)}, Dec 2013, pp. 1--6.

\bibitem{raps2}
M.~P. {Vestias} and H.~C. {Neto}, ``{Revisiting the Newton-Raphson Iterative
  Method for Decimal Division},'' in \emph{2011 21st International Conference
  on Field Programmable Logic and Applications}, Sep. 2011, pp. 138--143.

\bibitem{ScalingLessNewton}
E.~{Libessart}, M.~{Arzel}, C.~{Lahuec}, and F.~{Andriulli}, ``{A Scaling-Less
  Newton-Raphson Pipelined Implementation for a Fixed-Point Reciprocal
  Operator},'' \emph{IEEE Signal Processing Letters}, vol.~24, no.~6, pp.
  789--793, June 2017.

\bibitem{Gold1}
R.~E. Goldschmidt, \emph{{Applications of division by convergence}}, Cambridge,
  MA, USA, jun 1964.

\bibitem{Gold2}
M.~J. {Flynn}, ``{On Division by Functional Iteration},'' \emph{IEEE
  Transactions on Computers}, vol. C-19, no.~8, pp. 702--706, Aug 1970.

\bibitem{Goldschmidt}
M.~D. {Ercegovac}, L.~{Imbert}, D.~W. {Matula}, J.~. {Muller}, and G.~{Wei},
  ``{Improving Goldschmidt division, square root, and square root
  reciprocal},'' \emph{IEEE Transactions on Computers}, vol.~49, no.~7, pp.
  759--763, July 2000.

\bibitem{VLSICell}
V.~K. {Jain}, G.~E. {Perez}, and J.~M. {Wills}, ``{Novel reciprocal and
  square-root VLSI cell: architecture and application to signal processing},''
  in \emph{[Proceedings] ICASSP 91: 1991 International Conference on Acoustics,
  Speech, and Signal Processing}, April 1991, pp. 1201--1204 vol.2.

\bibitem{tab1}
H.~C. {Neto} and M.~P. {Vestias}, ``{Very low resource table-based FPGA
  evaluation of elementary functions},'' in \emph{2013 International Conference
  on Reconfigurable Computing and FPGAs (ReConFig)}, Dec 2013, pp. 1--6.

\bibitem{HighSpeedTC}
J.~{Pineiro} and J.~D. {Bruguera}, ``{High-speed double-precision computation
  of reciprocal, division, square root, and inverse square root},'' \emph{IEEE
  Transactions on Computers}, vol.~51, no.~12, pp. 1377--1388, Dec 2002.

\bibitem{ApproxOOX}
\BIBentryALTinterwordspacing
M.~Lunglmayr and O.~Ploder, ``{Fast approximate reciprocal approximations for
  iterative algorithms},'' 2020. [Online]. Available:
  \url{https://arxiv.org/abs/2007.06241}
\BIBentrySTDinterwordspacing

\bibitem{ElementaryMuller}
J.-M. Muller, \emph{{Elementary Functions - Algorithms and
  Implementation}}.\hskip 1em plus 0.5em minus 0.4em\relax Birkhäuser, 2016.

\bibitem{SFIXED}
\BIBentryALTinterwordspacing
D.~W. Bishop, ``{VHDL-2008 Support Library}.'' [Online]. Available:
  \url{https://github.com/FPHDL/fphdl}
\BIBentrySTDinterwordspacing

\end{thebibliography}
	
	%
	
	
	
	
	
	
	

\end{document}